\documentclass[aps,preprint,showpacs,showkeys,amsmath,amssymb,flexisym]{revtex4-1} 
\usepackage{graphicx}% Include figure files
\usepackage{dcolumn}% Align table columns on decimal point
\usepackage{bm}% bold math
\usepackage{hyperref}
\usepackage{color}
\usepackage{amssymb}
\begin{document}

%\preprint{APS/xxx-xxx}

\title{Neutrino phenomenology and scalar Dark Matter with $A_{4}$ flavor symmetry in Inverse and type II seesaw}
%\thanks{A footnote to the article title}
\author{Ananya Mukherjee}
\email{ananyam@tezu.ernet.in}
\author{Mrinal Kumar Das}
\email{mkdas@tezu.ernet.in}
\affiliation{%
 Department of Physics, Tezpur University, Tezpur 784\,028, India %
 }%

%\date{\today}

\begin{abstract}
We present a TeV scale seesaw mechanism for exploring the dark matter and neutrino phenomenology in the light of
recent neutrino and cosmology data. A different realization of the Inverse seesaw(ISS) mechanism with $A_{4}$ flavor symmetry is
being implemented as a leading contribution to the light neutrino mass matrix which usually gives rise to vanishing reactor mixing angle
 $\theta_{13}$. Using a non-diagonal form of Dirac neutrino mass matrix and $3\sigma$ values
 of mass square differences we parameterize the neutrino mass matrix in terms of Dirac Yukawa coupling ``$y$''. We then use type II seesaw
 as a perturbation which turns out to be active to have a non-vanishing reactor mixing angle without much disturbing the other neutrino oscillation
 parameters. Then we constrain a common parameter space satisfying the non-zero $\theta_{13}$,
 Yukawa coupling and the relic abundance of dark matter. Contributions of neutrinoless double beta decay are also included for standard as well as 
 non-standard interaction. This study may have relevance in future neutrino and Dark Matter experiments.
 
\end{abstract}

%Use showkeys class option if keyword
\maketitle

%\tableofcontents

\section{\label{sec:level1}Introduction}

The link between neutrino oscillation and modern cosmology needs an elucidation since both of them infer physics beyond 
Standard Model (BSM). Several theories have been deciphered to bridge between these two separate sectors of particle physics 
and cosmology \cite{Berlin15}. There is now a plethora of evidences for the existence of dark matter (DM) that constructs 
approximately a quarter of the energy density of the universe \cite{Beg91,Brad06,Benn13,Ade13}. Despite a number of recent
studies of simplified DM models their nature remains rather elusive. The most successful Standard Model (SM) of particle physics
too has not been able to furnish any signature of DM candidates and their properties. This is one of the pressing problems in
both high energy
physics and cosmology. This may surmise new physics beyond the standard model in near future. Therefore, searching for a concrete 
realization to provide a hint towards physics BSM will be of utmost interest. It will be more fascinating if the discovery 
of neutrino oscillation and the existence of DM can be framed within the same particle physics model.

 Presence of DM in the universe has been well established by astrophysics and cosmology experiments, although the exact particle 
 nature of DM is yet unknown. According to the Planck 2013 data \cite{Ade13}, 26.8\% of the energy density of the present 
 universe is composed of DM. The present abundance of DM or relic density is represented as 
  \begin{equation}\label{eq:1} 
  \Omega_{DM} h^2 = 0.1187\pm 0.0017,
   \end{equation}    
 Where,  $\Omega $ is the density parameter and h = Hubble parameter/100 is a parameter of order unity \cite{Dasgupta14}.

  Authors in \cite{marco08} proposed a ten-point test  that new particles have to pass
  in order to be considered as good DM candidates. The existence of dark matter is universally accepted, its nature remains elusive.
  It is usually assumed to be a single particle, but it may also be more than one. In specific
  models, it is often considered to be a fermion, scalar or vector \cite{Fraser14}. Among the requirements the
  potential DM candidate must meet, the stability is protected by invoking some parity symmetry like $Z_{2}$ which is 
  supposed to appear as a residual of a discrete flavor 
  symmetry. There have been extensive studies in this field adopting various flavor symmetry groups \cite{Valle10,Peinado11,Ma08}. 
  We have plenty of examples where different kinds of DM were extensively studied with their stability in 
  several ways. Recently connection between neutrino and the DM, using various flavor symmetries is drawing more attention in
  particle physics and cosmology. Here also we present a picture to construct a bridge between these two different sectors of particle
  physics adopting the $A_{4}$ based ISS realization. The most peculiar signatures of the ISS scenario are the additional decay channels of the Higgs boson
  into a heavy and ordinary neutrino, which confirms the SM particles to be a gateway to the scalar DM.
  In order for the SM particles being a portal to the dark sector, there must be at least
  two particles, one fermion and one 
  boson in the dark sector. Here in our model Higgs boson, is considered as a DM candidate, couples with SM neutrino through a right handed
  neutrino. Two neutral components of this Higgs which is a triplet under $A_{4}$ is responsible in making a correlation 
  with neutrino mass and Dark Matter. The stability of the 
  DM is explained by a remnant $Z_{2}$ symmetry. This $Z_{2}$ symmetry also 
  prevents the interaction of other particle contents of the model with the DM. Apart from the stability issue one more important test it must pass
  is to satisfy the observed relic density 
  given by equation (\ref{eq:1}). For getting the correct relic abundance we require to take the DM mass from 50 GeV 
  onwards. The Yukawa, which is responsible in making correlation between neutrino mass and DM coupling also needs to be fixed in such a way 
  that the potential DM candidate gives rise to correct relic abundance. 
  
  Several seesaw mechanisms have shown a promising role in explaining neutrino mass and mixing. The Inverse Seesaw (ISS) has been
  found to be an entirely different realization, which beautifully offers an explanation for having a tiny neutrino mass at the 
  cost of proposing the RH neutrino masses at the TeV scale which may be probed at the LHC experiments. The essence of the ISS lies in the
  fact that the double suppression by the mass scale associated with $M$ makes it possible to have such a scale much below than that 
  involved in the canonical seesaw mechanism. Which in turn renders us with SM neutrinos at sub-eV scale obtained with $m_{D}$ at 
  electroweak scale, $M$ at TeV scale and $\mu$ at keV scale as explained in \cite{Dias11}. This RH neutrino mass at TeV scale helps 
  us to get the required mediator mass in order to obtain the appropriate relic abundance of relics. In addition to the ISS we
  are working with the Type II seesaw mechanism which turns out to be instrumental to have the non-vanishing reactor mixing 
  angle. Both the inverse and type II seesaw are realized adopting the $A_{4}$ flavor symmetry. Then we have also studied the effective
  mass prediction to neutrinoless double beta decay for standard and non-standard contributions due to light neutrino exchanges. 

We organize the paper as follows. In Section.~\ref{sec:level2} we present our model.
Section.~\ref{sec:level3} provides the stability issue of DM. Non-zero reactor angle is explained in the  Section.~\ref{sec:level4}.
Section.~\ref{sec:level5} has been presented with the analysis on Neutrinoless double beta decay. Section.~\ref{sec:level6} offers the 
observation of the Relic abundance of DM in the context of the proposed model. In Section.~\ref{sec:level7} we have shown
the numerical analysis. Finally, in Section.~\ref{sec:level8} we end up with our conclusion.

\section{\label{sec:level2} Neutrino mass model with various seesaw scenarios}
\subsection{Inverse seesaw mechanism}
In our work we focus on the simplest ISS mechanism which is able to open up a new window to get the right handed neutrino mass
at a scale much below the one that involved in the canonical seesaw \cite{Dias11,Moha86,Mohapatra86,Dias12,Malin09,Ma09,Dev10,Dev12}. 

The fulfilment of the ISS scheme requires the extension of the SM fermion sector by the addition of three RH neutrinos 
$ N$ and three extra SM singlet neutral fermions $ S_{iL}$ to the active neutrinos $\nu_{iL}$ , with $ i = 1,2,3$ .
It is worth stating that, the implementation of the ISS allows us to make use of extra symmetries in order to provide the
neutrinos the following bilinear terms,

\begin{equation}
\mathcal{L} = - \bar{\nu}_{L}m_{D} N - \bar{S}_{L}M N - \frac{1}{2} \bar{S}_{L}\mu S^{C}_{L}+ H.C.,
\end{equation}

      The above Lagrangian implies a $ 9\times9 $ leptonic mass matrix,
      
      \begin{equation}\label{eq:3}
      M_{\nu}= \left(\begin{array}{ccc}
      0 & m_{D} & 0 \\
      m_{D}^{T}& 0 & M\\ 
      0 & M^{T} & \mu
      \end{array}\right).
      \end{equation}
      
      In spite of its many phenomenological successes the ISS has a drawback that the right-handed mass term in the $M_{\nu_{22}}$ entry of $M_{\nu}$ is allowed by
      symmetries. This is a typical problem of inverse seesaw models. But it is prevented here by using $Z_3$ symmetry.
       After block diagonalization of the equation (\ref{eq:3}) we get the lightest neutrino mass eigenvalue as ,
       
       \begin{equation}\label{eq:4} 
        m_{\nu}^{I} = m_{D}(M^{T})^{-1} \mu M^{-1}m_{D}^{T},
       \end{equation}

 which is considered as leading order contribution to the neutrino mass.
 Unlike the canonical seesaw mechanism that got its position in GUT, the ISS still lacks a special framework where the six 
 new neutrinos could find their places in the elemental particle content and normally can get a mass term. 
 
Non Abelian discrete flavor symmetries have played an important role in particle physics. In particular the symmetry 
 group $A_{4}$ have been immensely found of utmost operation \cite{Alt06,Alt10,Ma06,Brahm08,Oliver15}. In this work we have analyzed the model
presented by the authors in \cite{Valle10}, extended with additional flavons with inverse and type II seesaw. The flavor symmetry group $A_{4}$ is
the group of permutation of four objects, isomorphic to the symmetry group of a tetrahedron. $A_{4}$ has 
 four irreducible representations, among which there are three singlets and one triplet. 
 The group has got two generators. We summarize the $A_{4}$ based ISS model by assigning the matter
 fields as shown in Table.~\ref{tab1}. Four right handed neutrinos are introduced, three of which $N = (N_{1}, N_{2}, N_{3})$ are supposed to
 transform as a triplet of $A_{4}$ and the rest as a singlet $N_{4}$. We assign the SM 
 type Higgs $\eta$ to the $A_{4}$ triplet, which is considered as a 
 DM candidate in the present analysis. We have four additional SM fermion singlets among which `$S$' is transforming as 
  $A_{4}$ triplet and $S_{4}$ as $A_{4}$ singlet. To get a desired neutrino mass matrix structure 
 we are extending the Higgs sector by introducing six more Higgs fields, boosted by two additional symmetries $Z_{2}$ and $Z_{3}$ whose 
 quantum numbers are given in Table.~\ref{tab1}. The triplet multiplication rules of $A_{4}$ that has been used in the present
 analysis are given below (for more details see \cite{Ishi10,King13})

 The representations are given as follows
 \begin{gather*}
 a,b \sim 1,  \left(\begin{array}{c}
    a_{1}\\
    a_{2}\\
    a_{3}
   \end{array}\right) ,  
     \left(\begin{array}{c}
     {b_{1}}\\
    {b_{2}}\\
    {b_{3}}
   \end{array}\right) \sim 3.
 \end{gather*}
 
 \begin{gather*}
 (ab)_{1} = a_{1}b_{1} + a_{2}b_{2} + a_{3}b_{3}\\
 (ab)_{1^{\prime}} = a_{1}b_{1} + \omega a_{2}b_{2} + \omega^{2} a_{3}b_{3}\\
 (ab)_{1^{\prime\prime}} = a_{1}b_{1} +  \omega^{2} a_{2}b_{2} +  \omega a_{3}b_{3}\\
 (ab)_{3_{1}} = a_{2}b_{3} + a_{3}b_{1} + a_{1}b_{2}\\
 (ab)_{3_{2}} = a_{3}b_{2} + a_{1}b_{3} + a_{2}b_{1}
\end{gather*}

\subsection{Type II seesaw with triplet Higgs}     

For the type II seesaw mechanism to be implemented the SM is extended by the inclusion of an additional $SU(2)_{L}$ triplet
scalar field $\Delta$ having $U(1)_{Y}$ charge twice that of lepton doublets with its $2\times2$ matrix representation as
 
 \begin{equation} 
       \Delta = \left(\begin{array}{cc}
       \Delta^{+}/\sqrt{2} & \Delta^{++} \\
       \Delta^{0} & \Delta^{+}/\sqrt{2}
       \end{array}\right),
   \end{equation}

The Vacuum expectation value of the SM Higgs $<\phi_{0}>= v/\sqrt{2}$, the trilinear mass term $\mu_{\phi\Delta}$
generate an induced VEV for Higgs triplet as $\Delta^{0} = v_{\Delta}\sqrt{2}$ where, $v_{\Delta}\simeq\mu_{\phi\Delta}
v^{2}\sqrt{2}M^{2}_{\Delta}$ \cite{Borah14}.
 The type II seesaw contribution to light neutrino mass is given by
\begin{equation}\label{eq:9}
 m_{LL}^{II} = f_{\nu}v_{\Delta},
\end{equation}
 where the analytic formula for induced VEV  for neutral components of the Higgs scalar triplet, derived from the minimization 
 of the scalar potential \cite{Borah14}, is
 \begin{equation}
  v_{\Delta} \equiv \langle\Delta^{0}\rangle = \frac{\mu_{\phi\Delta}v^{2}}{\sqrt{2}M_{\Delta}^{2}}
 \end{equation}
 In the low scale type II seesaw mechanism operative at the TeV scale, barring the naturalness issue, one can consider a very small
 value of the trilinear mass parameter to be
 \begin{equation*}
  \mu_{\phi\Delta} \simeq 10^{-8} GeV.
 \end{equation*}
The sub-eV scale light neutrino mass with type II seesaw mechanism constrains the corresponding Majorana Yukawa
 coupling as 
 \begin{equation*}
  f_{\nu}^{2} < 1.4 \times 10^{-5}(\frac{M_{\Delta}}{1TeV})
 \end{equation*}

Within the reasonable value of $f_{\nu} \simeq 10^{-2}$ , the triplet Higgs scalar VEV is $v_{\Delta} \simeq 10^{-7} GeV$ which is in 
agreement with oscillation data. It is worth to note here that the tiny trilinear mass parameter $\mu_{\phi\Delta}$ controls the
neutrino overall mass scale, but does not play any role in the couplings with the fermions. The structure of the matrix $m_{LL}^{II}$,
with $w = f_{\nu} v_{\Delta}$ is explained in Section.~\ref{sec:level4}.

\section{\label{sec:level3}Stabilizing the Dark Matter}
A simple way to establish the stability of the DM is by invoking some parity symmetry like $Z_{2}$. Here is an attempt to 
search for a theory which is responsible for explaining neutrino phenomenology and Dark Matter (DM) stability as well. In this
ISS realization the symmetry $A_{4} \times Z_{2}\times Z_{3}$ spontaneously breaks to $Z_{2}$ accommodating a stable DM candidate.
The $A_{4}\times Z_{2}\times Z_{3}$
symmetry
here only allows the coupling of the $\eta$ with the singlet RH neutrinos rather than with charged fermions or quarks. It is worth noting that 
the alignment $\langle\eta\rangle \sim (1,0,0)$ breaks spontaneously $A_{4}\times Z_{2}\times Z_{3}$ to $Z_{2}$ since $(1,0,0)$ remains manifestly invariant
under one of the generators 
of the group $A_{4}$.

The stability of the DM candidate is guaranteed by this remnant symmetry. The $Z_{2}$ residual symmetry is defined by

\begin{gather*}
  N_{2}\rightarrow - N_{2} , S_{2}\rightarrow - S_{2}, \eta_{2}\rightarrow - \eta_{2}\\
   N_{3}\rightarrow - N_{3} , S_{3}\rightarrow - S_{3}, \eta_{3}\rightarrow - \eta_{3}\\
\end{gather*}

    The leading order Yukawa Lagrangian for the neutrino sector is given by the following equation.
    \begin{equation} \label{eq:8}
      \begin{split}
        \mathcal{L}^{I}_{\nu} = y_{1}^{\nu}L_{e}(N\eta)_{1} + y_{2}^{\nu}L_{\mu}(N\eta)_{1^{\prime\prime}} + y_{3}^{\nu}L_{\tau}(N\eta)_{1^{\prime}}+y_{4}^{\nu}L_{e}N_{4}h \\
       + y_{s}(SS)\phi_{s}+y_{s}^{\prime}S_{4}S_{4}\phi_{s}+y_{R}(NS)\phi_{R}+ y_{R}^{\prime}N_{4}S_{4}\phi_{R}. 
       \end{split}
       \end{equation}
        \begin{table}[htb]
        \centering
       \begin{tabular}{|c|c|c|c|c|c|c|c|c|c|c|c|c|c|c|c|c|c|}
       \hline & $ L_{e} $  & $ L_{\mu} $ & $ L_{\tau} $  & $l_{e}^{c}$ & $ l_{\mu}^{c}$  & $ l_{\tau}^{c} $ & $N$ & $N_{4}$ & $h$ & $\eta$ & $S_{4}$ & $S$ & $ \phi_{R} $ 
       & $\phi_{s} $  & $\zeta$ & $\xi$ & $\Delta$\\ 
       \hline $ SU(2)_{L} $ & 2  & 2 & 2 & 1 & 1  & 1 & 1  & 1 & 2 & 2 & 1 & 1 & 1 & 1 & 1 & 1 & 3\\ 
       \hline $ A_{4} $ & $ 1 $ &  $ 1^{\prime} $ & $ 1^{\prime\prime} $ & $ 1 $  & $ 1^{\prime\prime} $ & $ 1^{\prime} $ & $ 3 $ & $ 1 $ & $1$ & $3$ & $1$ &
       3 & 1 & 1 & $1^{\prime}$ & $1^{\prime\prime}$ & 1\\  
       \hline $ Z_{2} $ & 1 & 1 & 1& 1 & 1 & 1 & 1  &1 & 1 & 1 & -1 & -1 & -1 & 1 & 1 & 1 & 1 \\
       \hline $ Z_{3} $ & $\omega$ & $\omega$ & $\omega$ & $\omega^{2}$ & $\omega^{2}$ & $\omega^{2}$ & $\omega^{2}$ & $\omega^{2}$ & 1 & 1 & $\omega$ & $\omega$ & 1 &
       $\omega$ & 1 & 1 & $\omega$ \\
       \hline
       \end{tabular}
       \caption{Fields and their transformation properties under $ SU(2)_{L} $, the $ A_{4} $ flavor symmetry,
        $ Z_{2} $, $Z_{3}$ flavor symmetry} \label{tab1}  
      \end{table} 
    
     The following flavon alignments help us to get a desired neutrino mass matrix.
     
     $\langle\Phi_{R}\rangle = v_{R}$ , $\langle\Phi_{s}\rangle = v_{s}$ , $\langle h \rangle =  v_{h}$, $\langle \eta \rangle =  v_{\eta}(1,0,0)$.
     It is clear from the equation (\ref{eq:u},\ref{eq:v}) that, $m_{D}$ is connected to $v_{\eta}$ and $v_{h}$, and that $M$ is determined by the VEV $v_{R}$. In this way, the order
     of magnitude involved in the equation (\ref{eq:4}) is such that, $m_{\nu}\propto \frac{(v_{\eta} + v_{h})^{2}}{v_{R}^{2}}\mu$. Here $v_{\eta}$ and $v_{h}$ are
     of the order of
     electroweak breaking, $v_{R}$ is of the order of TeV scale. Therefore, to get $m_{\nu}$ in sub-eV, $\mu$ which is coming from the VEV of $\Phi_{S}$
     should be of the order of keV. The two components of $\eta$ are not generating the VEV \cite{Valle10}, considered potential
     DM candidate.
  Decomposition of the following terms present in the equation (\ref{eq:8}) has been shown as follows  
  \begin{gather*}
  y_{s}(SS)\phi_{s} =  y_{s}(S_{1}S_{1}+S_{2}S_{2}+S_{3}S_{3})\phi_{s},\\
   y_{R}(NS)\phi_{R} = y_{R}(N_{1}S_{1}+N_{2}S_{2}+N_{3}S_{3})\phi_{R}.
  \end{gather*}

    The chosen flavon alignments and the $A_{4}$ product rules allow us to have the Yukawa coupling matrices as follows 
      \begin{gather}\label{eq:u}
      m_{D}= \left(\begin{array}{cccc}
      y_{1}^{\nu}\langle\eta\rangle & 0 & 0 &y_{4}^{\nu}\langle h \rangle\\
      y_{2}^{\nu}\langle\eta\rangle& 0 & 0 & 0\\ 
      y_{3}^{\nu}\langle\eta\rangle & 0 & 0 & 0
      \end{array}\right)=\left(\begin{array}{cccc}
      x_{1}a & 0 & 0 & y_{1}b\\
      x_{2}a & 0 & 0 & 0\\ 
      x_{3}a & 0 & 0 & 0 
      \end{array}\right),
      \end{gather}
      \begin{gather}\label{eq:v}
      M= \left(\begin{array}{cccc}
      y_{R}\langle\phi_{R}\rangle & 0 & 0 & 0\\
      0 & y_{R}\langle\phi_{R}\rangle & 0 & 0\\ 
      0 & 0 & y_{R}\langle\phi_{R}\rangle & 0\\
      0 & 0 & 0 & y_{R}^{\prime}\langle\phi_{R}\rangle
      \end{array}\right) = \left(\begin{array}{cccc}
      M_{1} & 0 & 0 & 0\\
      0 & M_{1} & 0 & 0\\ 
      0 & 0 & M_{1} & 0\\
      0 & 0 & 0 & M_{2}
      \end{array}\right),
      \end{gather}

      \begin{equation}\label{eq:w}
     \mu_{s}= \left(\begin{array}{cccc}
      y_{s}\langle\phi_{s}\rangle & 0 & 0 & 0\\
      0& y_{s}\langle\phi_{s}\rangle & 0 & 0\\ 
      0 & 0 & y_{s}\langle\phi_{s}\rangle & 0\\
      0& 0 & 0 & y_{s}^{\prime}\langle\phi_{s}\rangle
      \end{array}\right)= \left(\begin{array}{cccc}
      \mu_{1} & 0 & 0 & 0\\
      0 & \mu_{1} & 0 & 0\\ 
      0 & 0 & \mu_{1} & 0\\
      0 & 0 & 0 & \mu_{2}
      \end{array}\right),
      \end{equation}

    The above three matrices lead to the following light neutrino mass matrix under the ISS framework
    \begin{equation}\label{eq:b}
     m_{\nu} = \left(\begin{array}{ccc}
      \frac{y_{1}^{2}b^{2}\mu_{2}}{M_{2}^{2}}+ \frac{a^{2}x_{1}^{2}\mu_{1}}{M_{1}^{2}} & \frac
      {a^{2}x_{1}x_{2}\mu_{1}}{M_{1}^{2}} & \frac{a^{2}x_{1}x_{3}\mu_{1}}{M_{1}^{2}}\\
      \frac{a^{2}x_{1}x_{2}\mu_{1}}{M_{1}^{2}} & \frac{a^{2}x_{2}^{2}\mu_{1}}{M_{1}^{2}} & \frac{a^{2}x_{2}x_{3}\mu_{1}}{M_{1}^{2}} \\ 
      \frac{a^{2}x_{1}x_{3}\mu_{1}}{M_{1}^{2}} & \frac{a^{2}x_{2}x_{3}\mu_{1}}{M_{1}^{2}} & \frac{a^{2}x_{3}^{2}\mu_{1}}{M_{1}^{2}} 
      \end{array}\right).
    \end{equation}
          
 The assigned $A_{4}$ charge of this Higgs triplet $\eta$ restricts the interaction of $\eta$ with the charged leptons. Now the 
  Lagrangian for the charged lepton mass is given by,
   
   \begin{equation}
     \mathcal{L}^{I}_{l} = y_{e}L_{e}l_{e}^{c}h + y_{\mu}L_{\mu}l_{\mu}^{c}h + y_{\tau}L_{\tau}l_{\tau}^{c}h,
     \end{equation}
     
 Following is the mass matrix for charged leptons.    
   \begin{equation}
   m_{l} = \left(\begin{array}{ccc}
     y_{e}\langle h \rangle & 0 & 0\\
      0 & y_{\mu}\langle h \rangle & 0\\ 
      0 & 0  & y_{\tau}\langle h \rangle 
      \end{array}\right).
    \end{equation}
\section{\label{sec:level4}The reactor mixing angle}

It is needless to say that there is a menagerie of theories, put forward in establishing the $\theta_{13}$ as having a 
nonzero value. Here also we are trying to present such a picture by including a perturbation called type II perturbation
to the above Lagrangian given by equation(\ref{eq:8}) which is realized within the type II seesaw mechanism \cite{Borah14,Ma98,Rod04,Lin07,Borahd13,Borahd14}. 
The type II seesaw term is followed by this term
 
  %   \mathcal{L}^{II} = f_{\nu}\frac{LL\zeta\triangle}{\Lambda} + f_{\nu}\frac{LL\xi\triangle}{\Lambda}, 
   \begin{equation}
   \mathcal{L}^{II} = f_{\nu}\frac{(L_{e}L_{\tau}+L_{\mu}L_{\mu}+L_{\tau}L_{e})\zeta\Delta}{\Lambda}  +
   f_{\nu}\frac{(L_{e}L_{\mu}+L_{\mu}L_{e}+L_{\tau}L_{\tau})\xi\Delta}{\Lambda},
    \end{equation}
    
Where, $\Lambda$ is the cutoff scale. With the type II perturbation the Lagrangian takes the following form

 \begin{equation} 
 \begin{split}
  \mathcal{L} = y_{e}L_{e}l_{e}^{c}h + y_{\mu}L_{\mu}l_{\mu}^{c}h +y_{\tau}L_{\tau}l_{\tau}^{c}h + y_{1}^{\nu}L_{e}(N\eta)_{1} + y_{2}^{\nu}L_{\mu}(N\eta)_{1^{\prime\prime}}+y_{3}^{\nu}L_{\tau}(N\eta)_{1^{\prime}} \\
  +y_{4}^{\nu}L_{e}N_{4}h +y_{s}(SS)\phi_{s}+y_{s}^{\prime}S_{4}S_{4}\phi_{s}+y_{R}(NS)\phi_{R}+y_{R}^{\prime}N_{4}S_{4}\phi_{R}\\
   + f_{\nu}\frac{(L_{e}L_{\tau}+ L_{\mu}L_{\mu}+L_{\tau}L_{e})\zeta\Delta}{\Lambda} +f_{\nu}\frac{(L_{e}L_{\mu}+L_{\mu}L_{e}+L_{\tau}L_{\tau})\xi\Delta}{\Lambda}.
   \end{split}
 \end{equation}
 The last two terms represent the perturbation to the leading order terms in the above Lagrangian giving rise to non-zero $\theta_{13}$.    

 Here we have implemented the $A_{4}$ group to explain the structure of the neutrino mass matrix 
 (\ref{eq:17}) originating from the type II seesaw mechanism. The $SU(2)_{L}$ triplet Higgs field $\Delta_{L}$ is supposed to
transform as a $A_{4}$ singlet. Two more flavon fields $\zeta$ and $\xi$ have been introduced which are assumed to
 transform as $A_{4}$ singlets as summarized in the Table.~\ref{tab1}. The flavon alignments which help in constructing the $m_{LL}^{II}$ matrix
 are as follows

 $\langle \Delta \rangle \sim v_{\Delta}$, $\langle \zeta \rangle\sim v_{\zeta}$, $ \langle \xi \rangle \sim v_{\xi}$.
 $\zeta$ and $\xi$ are assumed to take the VEV in the same scale $v_{\zeta} = v_{\xi} = \Lambda $. With these flavon alignments the structure
  of mass matrix $m^{II}_{LL}$ will take the form 
  \begin{equation}\label{eq:17}
   m_{LL}^{II} = \left(\begin{array}{ccc}
   0 & -w & w \\
  -w & w & 0\\ 
   w & 0 & -w
   \end{array}\right).
   \end{equation}

 \section{\label{sec:level5}  Neutrinoless Double Beta Decay}       
The time period for Neutrinoless Double Beta Decay rate is
directly proportional to the square of the effective neutrino mass $m^{ee}_{\nu}$. Which implies that in determining the time period
for Neutrinoless Double Beta Decay, the effective mass plays a non-trivial role in the standard three generation picture.
The effective neutrino mass can be given by
\begin{equation}
 |m^{ee}_{\nu}| = |U^{2}_{ei} m_{i}|,
\end{equation}

\begin{figure}[h]
\centering
\includegraphics[width=0.4\textwidth]{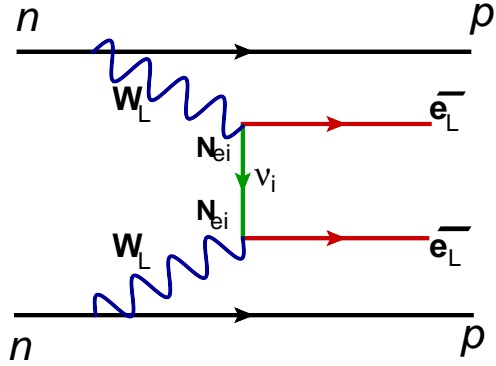}
\caption{Feynman diagram contributing to neutrinoless double beta decay due to light
 neutrino exchanges.}
\label{fig1}
\end{figure}

The Unitary matrix is the PMNS matrix which is the neutrino mixing matrix in the basis where the charged lepton mass 
matrix is diagonal \cite{Chakra12,Awasthi13}. In addition to this, following non-standard contributions become transparent
in the present model. 
\begin{itemize}
 \item Two separate contributions due to light and heavy neutrino exchanges to $0\nu\beta\beta$ 
come into play. And this event is established
by writing the flavor eigenstates as a linear combination of light and heavy mass eigenstates. The only contribution that becomes effective
in the ISS regime comes from the contribution due to light neutrino exchanges. 

\begin{equation}
 \nu_{\alpha} = N_{\alpha i}\nu_{i} + U_{\alpha j} \xi_{j},
\end{equation}
where, $N_{\alpha i}$ and $U_{\alpha j}$ are the mixing matrices for light and heavy neutrino respectively.
The effective mass takes different values depending on the framework (quasi degenerate or normal/inverted 
hierarchies), the neutrino mass states are in. Now considering the light neutrino contribution (the only contribution to ISS in this model),
the key formula for determining the effective neutrino mass is given by
\begin{equation}
 m^{ee}_{\nu,LL} \simeq U^{2}_{e1} m_{1} + U^{2}_{e2} e^{2i\alpha} m_{2} + U^{2}_{e3} e^{2i\beta} m_{3}.
\end{equation}
\item The triplet Higgs contribution from the type II seesaw. The contribution from the triplet Higgs is of the
order of $10^{-13} m_{i}$ which is much suppressed as compared to the dominant contributions \cite{Chakra12}. 
\end{itemize}

  Of special importance is the fact that, the chosen value of Yukawa coupling giving rise to the observed relic abundance of our DM candidate, constrains
  the lightest neutrino mass significantly in the presented forum. 
  The fine tuned Yukawa couplings $(0.994 - 1)$ is noticed to play an important role in
  achieving the lightest neutrino mass and in turn to get the effective neutrino mass prediction within the GERDA bound $(0.5eV)$. The type II perturbation strength
  is found to play some role in giving $m_{lightest}$ within the PLANK bound ($0.065$ eV for IH). The introduced model also evinces
  the role of leptonic mixing matrix elements and the lightest neutrino mass as the effective neutrino mass is dependent upon them.

  \section{\label{sec:level6} Relic Density of Dark Matter}     
\begin{figure}[h]
\centering
\includegraphics[width=0.4\textwidth]{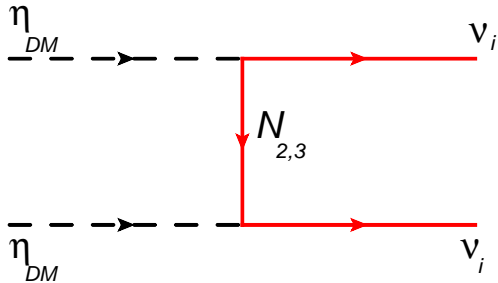}
\caption{Feynman diagram showing the scattering of $\eta_{2}$ and $\eta_{3}$.}
\label{fig2}
\end{figure}

The relic abundance of a DM particle $\chi$ is given by the Boltzmann equation \cite{Griest91,Kolb90,Gon97,Gel91}
    
    \begin{equation}
    \frac{dn_{\chi}}{dt} + 3Hn_{\chi} = - <\sigma v> (n^{2}_{\chi}- (n^{eqb}_{\chi})^{2}),
    \end{equation}
    
where $ n_{\chi} $ is the number density of the DM particle $ \chi $ and $ n^{eqb}_{\chi} $ is the number density when 
    $ \chi $ was in thermal equilibrium. $H$ is the Hubble rate and $<\sigma v>$ is the thermally averaged annihilation
    cross-section of the DM particle $ \chi $. Numerical solution of the Boltzmann equation is given by \cite{Kolb90}
     \begin{equation}
    \Omega_{\chi}h^{2} \approx \frac{1.04 \times 10^{9}x_{F}}{M_{pl}\sqrt{g_{*}}(a + 3b/x_{F})},
    \end{equation}
    where $x_{F}= \frac{m_{\chi}}{T_{F}}$ , $T_{F}$ is the freeze-out temperature, $g_{*}$ is the number of relativistic 
    degrees of freedom at the time of freeze-out. DM particles with electroweak scale mass and couplings freeze out at
    temperatures in the range $x_{F}\approx 20-30$. This in turn simplifies to \cite{Jung96}
    \begin{equation}
    \Omega_{\chi}h^{2} \approx \frac{3 \times 10^{-27} cm^{3}s^{-1}}{<\sigma v>}.
    \end{equation}
    For complex scalar DM, the annihilation rate is given by equation (\ref{eq:d}). The relic abundance is related to the 
    cross section of the DM-DM interaction. The terms in equation (\ref{eq:8}) evinces the interaction \ref{fig2}. While 
    finding the allowed parameter space satisfying the correct relic abundance 
    and neutrino oscillation parameters we vary the Relic mass and the Majorana fermion mass (the right handed neutrino) both of
    which are involved in the cross section formula as shown in \cite{Bai13} reads as
     \begin{equation}\label{eq:d}
     (\sigma v )^{\chi\chi\dagger}_{complex scalar} = \frac{v^{2} y^{4} m ^{2}_{\chi}}{48\pi(m ^{2}_{\chi} + m ^{2}_{\psi})^{2}}. 
     \end{equation} 
 With $v$ = relative velocity of the two relic particles and is typically $0.3c$ at the freeze out temperature, $ \chi $ is the 
 relic particle (DM), $y$ is the Yukawa coupling, $ m_{\chi} $ the mass of the relic, $ m_{\psi} $ is the mass of 
 the right handed neutrino.
 
\begin{figure}[h]
\centering
\includegraphics[width=0.6\textwidth]{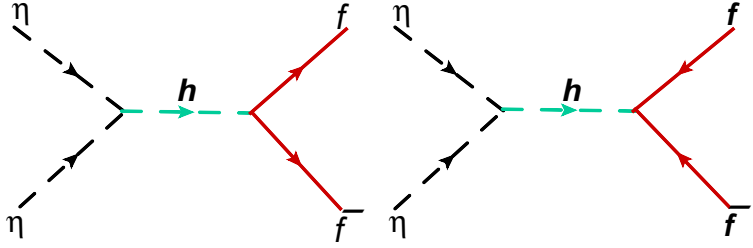}
\caption{Self annihilation of $\eta_{2}$ and $\eta_{3}$ into SM fermions(conventions are followed from \cite{Dreiner10}).}
\label{fig3}
\end{figure}

 The dark matter relic abundance may get affected by some kind of annihilation processes which might have taken place between the two neutral scalars
 depending on their mass difference $\Delta m = m_{\eta_{2}} - m_{\eta_{3}}$. If the mass splitting is of the order of freeze-out temperature, $T_{f}$ 
 the coannihilation between the two neutral scalars play a significant role in finding the relic abundance of dark matter. But if $\Delta m$ is larger than
 the freeze-out temperature, then the immediate heavier neutral scalar affects the dark matter relic density notably.
 The self annihilation between dark matter and next to lightest neutral component of scalar triplet $\eta$ contribute to the annihilation cross
 section of dark matter. Many authors in \cite{Griest91,Gon97,Bell14} explored this kind of self annihilation effects on dark matter relic abundance.
 To calculate the effective annihilation cross section we are following the analysis of \cite{Griest91}. The various
 annihilation channels and interactions can be given by figure \ref{fig3}.
 
 For low mass scheme ($m_{DM} < M_{W}$), the self annihilation of either $\eta_{2} $ or $\eta_{3}$ into SM particles takes place via SM Higgs boson
  as shown in figure \ref{fig3}. The according annihilation cross section \cite{Gon97,Bell14} is followed by equation (\ref{eq:z}).
  
  \begin{equation}\label{eq:z}
   \sigma_{xx} = \frac{|Y_{f}|^2|\lambda_{x}|^2}{16\pi s} \frac{(s - 4m_{f}^2)^{3/2}}{\sqrt{s-4m_{x}^{2}}
   ((s - m_{h}^{2})^2 + m_{h}^{2}\Gamma_{h}^{2})} , 
  \end{equation}

  where $x\rightarrow \eta_{2,3}$, $\lambda_{x}$ is the coupling of $x$ with SM Higgs boson $h$ and $Y_{f}$ is the Yukawa coupling of fermions,
  which has been estimated to be $0.32$ albeit the full possible range of values is $\lambda_{f}= 0.26-0.63$ \cite{Dasgupta14}.
  $\Gamma_{h} = 4.15 MeV$ is the SM Higgs decay width, $m_{h}$ is 126 GeV. $s$ is the thermally averaged center of mass squared energy given by
   \begin{equation}
    s = 4m^{2} + m^{2}v^{2}.
  \end{equation}
  
where, $v$ is the relative velocity and $m$ is the mass of the relic. 
 In order to yield the correct relic abundance we need to constrain the Yukawa coupling along with the relic mass and the 
 mediator mass. Similar to the works done in \cite{Bouc11, Bouc12} here also we consider the neutral component of the scalar 
 triplet as the DM candidate. We choose the relic mass as lighter than the W boson mass $ m_{DM} \le M_{W} $. And interestingly 
 for the relic we stick to a comparatively low mass region, which is around 50 GeV. The mediator mass here in our case, i.e., 
 the Majorana neutrino mass is required to vary from 153 GeV to 154 Gev to obtain the observed relic density. This type of findings
 have been extensively studied in the literature \cite{Bai13,Bai14}. For a light DM with a mass below 10 GeV, the LHC searches have a better
 awareness for complex scalar DM cases. Moreover, the LHC has a better reach than direct 
 detection experiments with DM masses up to around 500 GeV for the complex scalar DM case. 
  
\section{\label{sec:level7} Numerical Analysis}
   
 The latest global fit \cite{Berg15} value with their best fit point (bfp) for $3\sigma $ range of neutrino oscillation
 parameters used to study neutrino phenomenology are given in Table.~\ref{tab2} and Table.~\ref{tab3}:
 
   \begin{table}[htb]
        \centering
       \begin{tabular}{|c|c|c|}
       \hline  Oscillation parameters   & bfp & $3\sigma$ Cl \\ 
       \hline $ \Delta m^{2}_{21} [10^{-5} eV^{2}] $ & $7.5$  & $(7.02 , 8.07)$   \\ 
       \hline $ \Delta m^{2}_{31}[10^{-3} eV^{2}] $ & $ 2.457 $ & $(2.317 , 2.607)$  \\ 
       \hline $  \sin ^{2}\theta_{12} $ & $0.304$ & $(0.270 , 0.344)$    \\ 
       \hline $ \sin ^{2}\theta_{13} $ & $0.0218$ &  $(0.0186 , 0.0250)$  \\ 
       \hline  $\sin ^{2}\theta_{23}$ & -- & 0.381-0.643   \\
       \hline
       \end{tabular}
     \caption{Neutrino Oscillation data for Normal mass Ordering} \label{tab2}       
           \end{table} 
        
     \begin{table}[htb]
        \centering
       \begin{tabular}{|c|c|c|}
       \hline  Oscillation parameters   & bfp & $3\sigma$ Cl \\ 
       \hline $ \Delta m^{2}_{21} [10^{-5} eV^{2}] $ & $7.5$  & $(7.02 , 8.07)$   \\ 
       \hline $ \Delta m^{2}_{23}[10^{-3} eV^{2}] $ & $-2.449$ &  $-2.590 , -2.307$ \\ 
       \hline  $\sin ^{2}\theta_{12}$  &  $0.304$  &  $0.270 , 0.34$   \\ 
       \hline  $\sin ^{2}\theta_{13}$  & $0.0219$  &  $0.0188 , 0.0251$  \\ 
       \hline  $\sin ^{2}\theta_{23}$ & -- & $0.388 , 0.644$ \\
       \hline
       \end{tabular}
      \caption{Neutrino Oscillation data for Inverted mass Ordering} \label{tab3}       
        \end{table}     
       
 Cosmological constraint says that,
\begin{equation*}
 m_{1}+ m_{2}+m_{3} \le 0.23  eV.
\end{equation*}  
      
\begin{figure*}[h]
\centering
\includegraphics[width=0.9\textwidth]{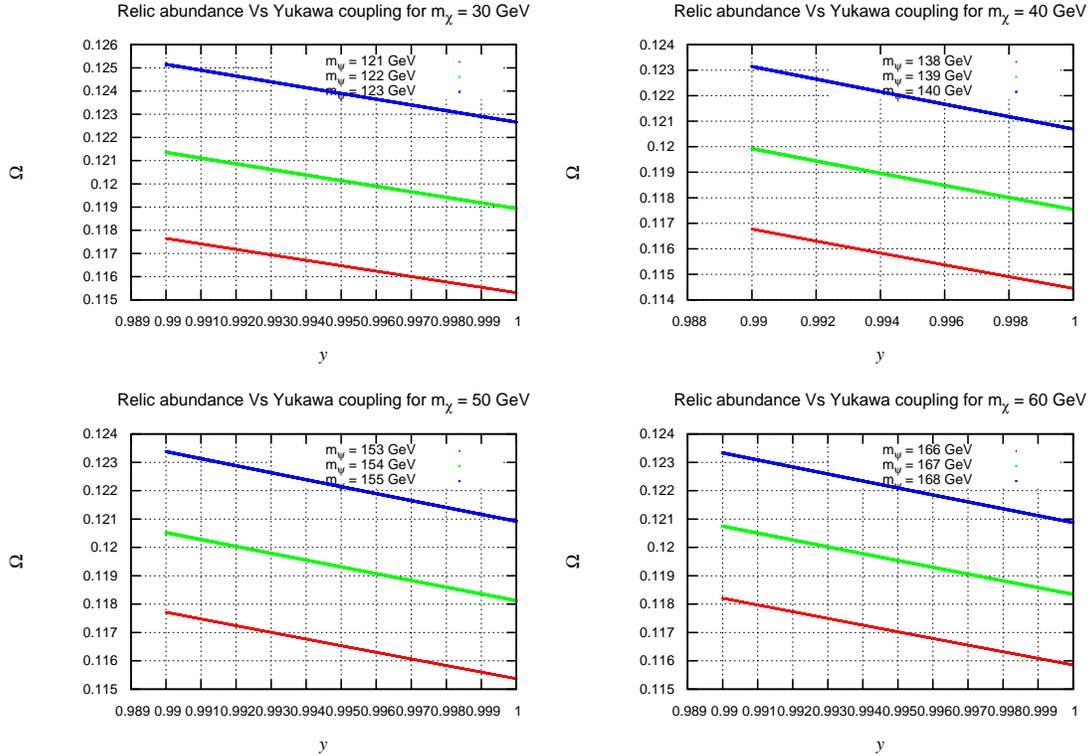}
\caption{Variation of relic abundance with Yukawa coupling.}
\label{fig4}
\end{figure*}

 The Yukawa coupling governing the interaction is present in the established mathematical expression which
 computes the scattering
 cross section of this interaction in turn the relic abundance of the potential DM. As a proper choice of Yukawa coupling, 
 the mediator mass along with the complex scalar mass allows us to achieve the observed
 relic abundance we need to put constraints on them. In our work we first fix the above mentioned parameters to get the relic 
 abundance which is reported by  PLANCK 2013 data. 
 Fixing the relic mass around 50 GeV and varying the mediator mass from 153 to 154 GeV we get the idea of 
Yukawa coupling yielding the correct relic abundance. Since the required relic abundance for the potential DM candidate
desires a mediator mass at a much lower scale (around 153 GeV) the ISS realization helps us to keep the RH neutrino (which
is here, the mediator particle governing the t-channel scattering as shown in \ref{fig2}) mass at a scale much below than that one involved in the
canonical seesaw. The Yukawa coupling needs to fall between 0.99 to 1 to have a better reach of the relic abundance as shown in
 figure \ref{fig4}. 
   
We redefine the parameters of the matrix shown by the equation(\ref{eq:b})in terms of $p$, $q$ and $r$. Where, $p= \frac{ax_{1}\sqrt{\mu_{1}}}{M_{1}}$,
 $q= \frac{ax_{2}\sqrt{\mu_{1}}}{M_{1}}$ and $r= \frac{ax_{3}\sqrt{\mu_{1}}}{M_{1}}$. From the 
 requirement of bringing the light neutrino mass matrix into TBM form, we equate the 11-element of $m_{\nu}$ to $2q^{2}-pq$ \cite{Valle10}.
 This is done in accordance with adjusting the Yukawa couplings and the associated VEVs.
 Along with this redefinition we also make $q=r$ by $x_{2}=x_{3}$ for numerical analysis. This form of light neutrino mass matrix has an 
   inverse hierarchial neutrino mass spectrum and a zero eigenvalue with $m_{3}=0$. For numerical analysis, we take another couple of definitions
   for the Yukawa couplings $x_{1} = x$ and $x_{2} = x_{3}= y$.
   We have kept $x = 1$ and varied $y$ for computing the oscillation parameters and 
$m^{ee}_{\nu}$, however, there is no significant changes observed by keeping $y$ fixed and varying $x$.
Each value of $y$ gives rise to various sets of the neutrino mass matrix parameters $p,q$. We parameterize the light neutrino mass matrix
obtained from the ISS realisation with the help of recent neutrino oscillation data given 
in Table.~\ref{tab2} and Table.~\ref{tab3}.
Along with the redefined parameters of the light neutrino mass matrix
    and using equation (\ref{eq:u}), (\ref{eq:v}), (\ref{eq:w})
 the new light neutrino mass matrix is found to be of TBM type given by equation(\ref{eq:q})

   \begin{equation}\label{eq:q}
    m_{\nu} = \left(\begin{array}{ccc}
    2q^{2} - pq & pq & pq \\
    pq & q^{2} & q^{2}\\ 
    pq & q^{2} & q^{2}
    \end{array}\right).
  \end{equation}
  
\begin{figure*}
\centering
\includegraphics[width=10cm,height=8cm]{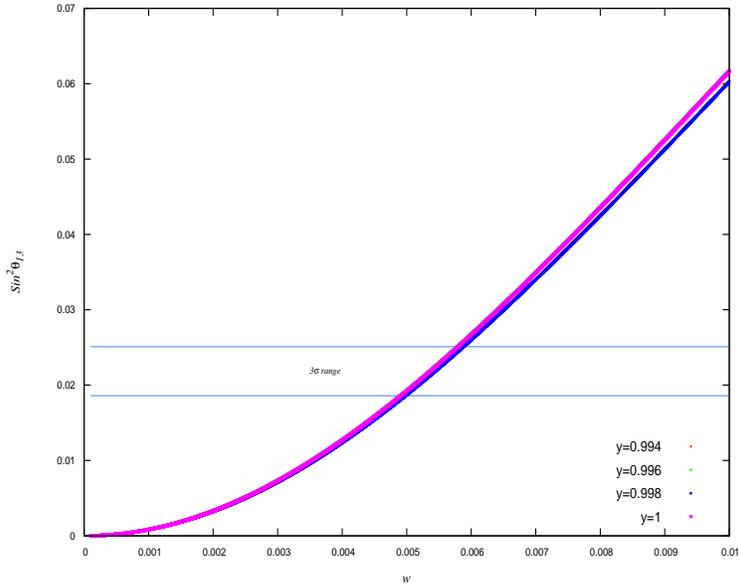}
\caption{Generation of non-zero $sin^2\theta_{13}$ varying the type II strength for best fit values.}
\label{fig5}
\end{figure*}

We have analyzed the model only for the IH case as the light neutrino mass matrix structure only allows us to have the inverted hierarchy mass pattern. 
  After diagonalizing the complete mass matrix the mass eigenvalues are found to be $m_{1} = -2(pq - q^2)$,
   $m_{2} = q(p+2q)$ and $m_{3} = 0$. Then we parameterize the mass matrix keeping $x=1$ while
  at the same time varying $y$ between a range around $0.994 - 1$. Choosing each set of $p,q$ values which have been found different for different ``$y$''
  values, we get several light neutrino mass matrices. The same Yukawa coupling $y$ is being varied in the dark matter sector too for showing
  its contribution to obtain the correct relic abundance. With the discovery of non-zero reactor mixing angle, it is a customary to reflect the concept theoretically. In our work 
  we try to provide a platform which reproduces the same. For that purpose, we include type II perturbation \cite{Borah14} to
  the leading order neutrino mass matrix as explained in Section.~\ref{sec:level4}. This perturbation brings out non-zero $\theta_{13}$ in $3\sigma$
  range along with $m_{3}\neq 0$ leaving the light neutrino masses with IH nature only. The numerical value of the perturbation 
  term $w = f_{\nu}v_{\Delta}$ critically depends upon the Majorana coupling $f_{\nu}$, trilinear mass parameter $\mu\phi\Delta$,
  and $M$. Accordingly, we vary the type II seesaw strength from $10^{-6}$ to $0.01$ to produce non-zero $\theta_{13}$. 
  It is observed from the figure \ref{fig5} that, the type
  II seesaw strength of $10^{-3}$ eV is generating the non-zero $\theta_{13}$ in the $3\sigma$ range in all cases.

  The perturbation matrix takes the following structure 
  \begin{equation*}
     m^{II}_{\nu} = \left(\begin{array}{ccc}
     0 & -w & w \\
    -w & w & 0\\ 
    w & 0 & -w
    \end{array}\right),
   \end{equation*}
   
     After adding the perturbation we get the neutrino mass matrix as follows
     \begin{equation*}
      m_{\nu} = m_{\nu}^{I} + m_{\nu}^{II}.
     \end{equation*}
     
 %\begin{equation}
 %   m_{\nu} = \left(\begin{array}{ccc}
 %   2q^{2} - pq & pq- w & pq + w \\
 %   pq- w & q^{2}+ w & q^{2}\\ 
 %   pq + w & q^{2} & q^{2}- w
 %   \end{array}\right)
 % \end{equation}
   Now the elements of these diagonalized matrices are associated with the parameters of the model and the type II perturbation term. The set of $p,q$ values 
   obtained for each $y$ value and chosen for analysis are listed in Table.~\ref{tab4}, Table.~\ref{tab5}, Table.~\ref{tab6}. In addition
   $p,q$ corresponds to some complex sets of solution too.Taking them under consideration, no significant changes in the numerical analysis have been observed.

 A comparison among the various sets of results obtained in the DM phenomenology part has been made in table\ref{tab10} and 
neutrino phenomenology has been shown in the table \ref{tab11}.

   \begin{table}[htb]
     \centering
     \begin{tabular}{|c|c|c|c|c|}
  \hline   Parameters & $y=0.994$ & $y=0.996$ & $y=0.998$ & $y=1$   \\ 
    \hline $p$ &  $0.366138$ & $0.366146$ & $0.366154$ & $0.357719$\\ 
    \hline $q$ &   $0.0899502$ & $0.089768$ & $0.0895865$ & $0.091516$\\  
    \hline
    
     \end{tabular}  
     
     \caption{Values of $p,q$ obtained by solving for IH case with best fit central value of $3\sigma$ Deviations} \label{tab4}
      
     \end{table}

   \begin{table}[htb]
    \centering
     \begin{tabular}{|c|c|c|c|c|}
    \hline    Parameters &   $y=0.994$ & $y=0.996$ & $y=0.998$ & $y=1$  \\ 
    \hline $p$ &  $0.371351$ & $0.371359$ & $0.371367$ & $0.362663$ \\ 
     \hline $q$ & $ 0.0911924$ & $0.0910077$ & $0.0908236$ & $0.0928181$\\ 
      \hline
    
      \end{tabular}  
      
      \caption{Values of $p,q$ obtained by solving for IH case with a upper bound of $3\sigma$ Deviations} \label{tab5}
      
     \end{table}

 \begin{table*}[htb]
 \centering
     \begin{tabular}{|c|c|c|c|c|}
      \hline    Parameters &   $y=0.994$ & $y=0.996$ & $y=0.998$ & $y=1$  \\ 
    \hline $p$ &  $0.360693$ & $0.3607$ & $0.360708$ & $0.352452$ \\ 
     \hline $q$ & $ 0.088626$ & $0.0884465$ & $0.0882677$ & $0.0901551$\\ 
      \hline
       \end{tabular}  
       \caption{Values of $p,q$ obtained by solving for IH case with an lower bound of $3\sigma$ Deviations} \label{tab6}     
     \end{table*}

 The light neutrino mass matrix (\ref{eq:q}) is having only two unknown parameters, solution for which demands two equations. 
 Two masses squared differences 
 which we get from neutrino oscillation datas, lead to those two parameters. Then, using the solutions for $p$ and $q$ 
 the light neutrino mass matrix is obtained. Then we fix the mass eigenvalues from that light neutrino mass matrix.
 
  Using the best fit central values from the oscillation data, we numerically fit the leading order neutrino mass matrix. A 
  thorough analysis has been carried out to check whether the oscillation parameters are near to reach or not by taking the 
  upper and lower bound of $3\sigma$ deviation as well. 
  Here we try to exhibit an unexplored parameter space satisfying both the DM relic abundance and 
 neutrino phenomenology.
 
\begin{figure*}
 \centering
 \includegraphics[width=10cm,height=8cm]{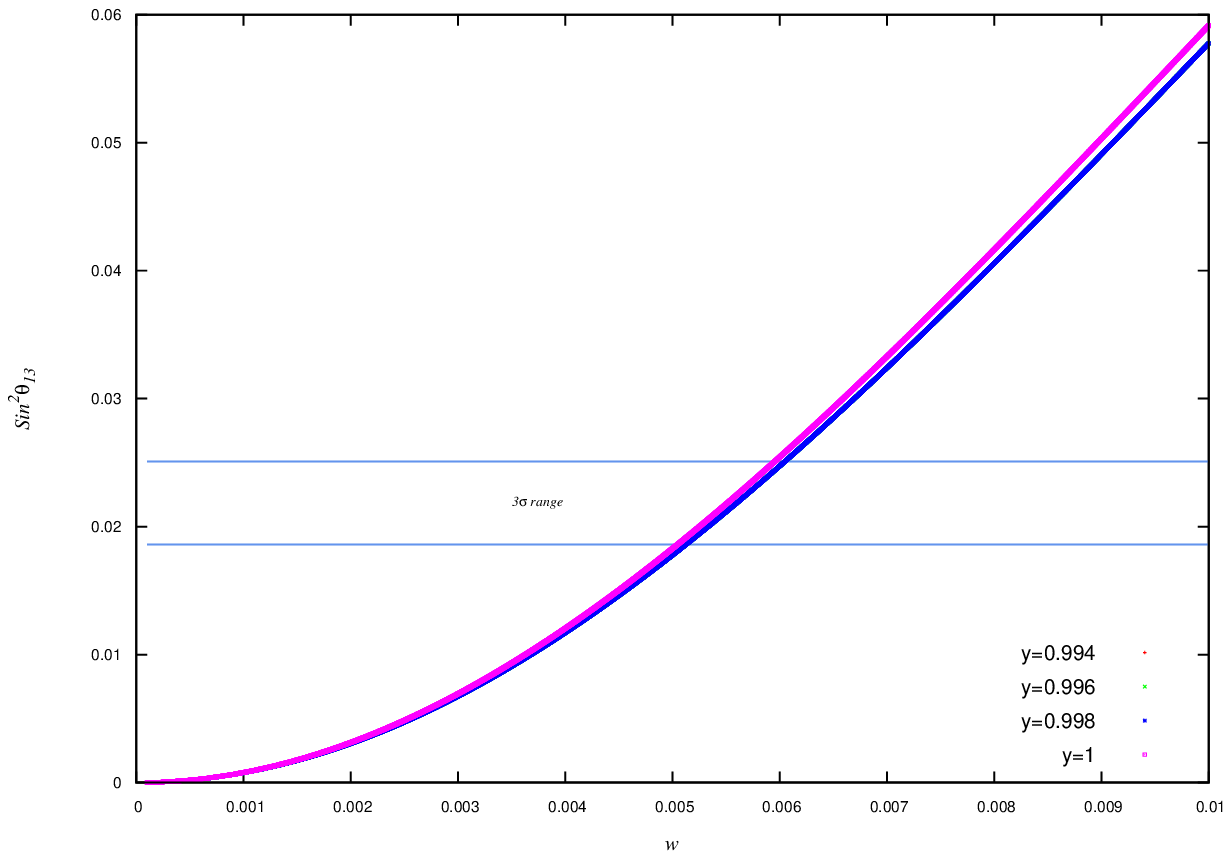}
 \caption{Generation of non-zero $sin^2\theta_{13}$, varying the type II strength using upper bound of $3\sigma$ deviations. }
 \label{fig6}
 \end{figure*}

\begin{figure*}
 \centering
 \includegraphics[width=10cm,height=8cm]{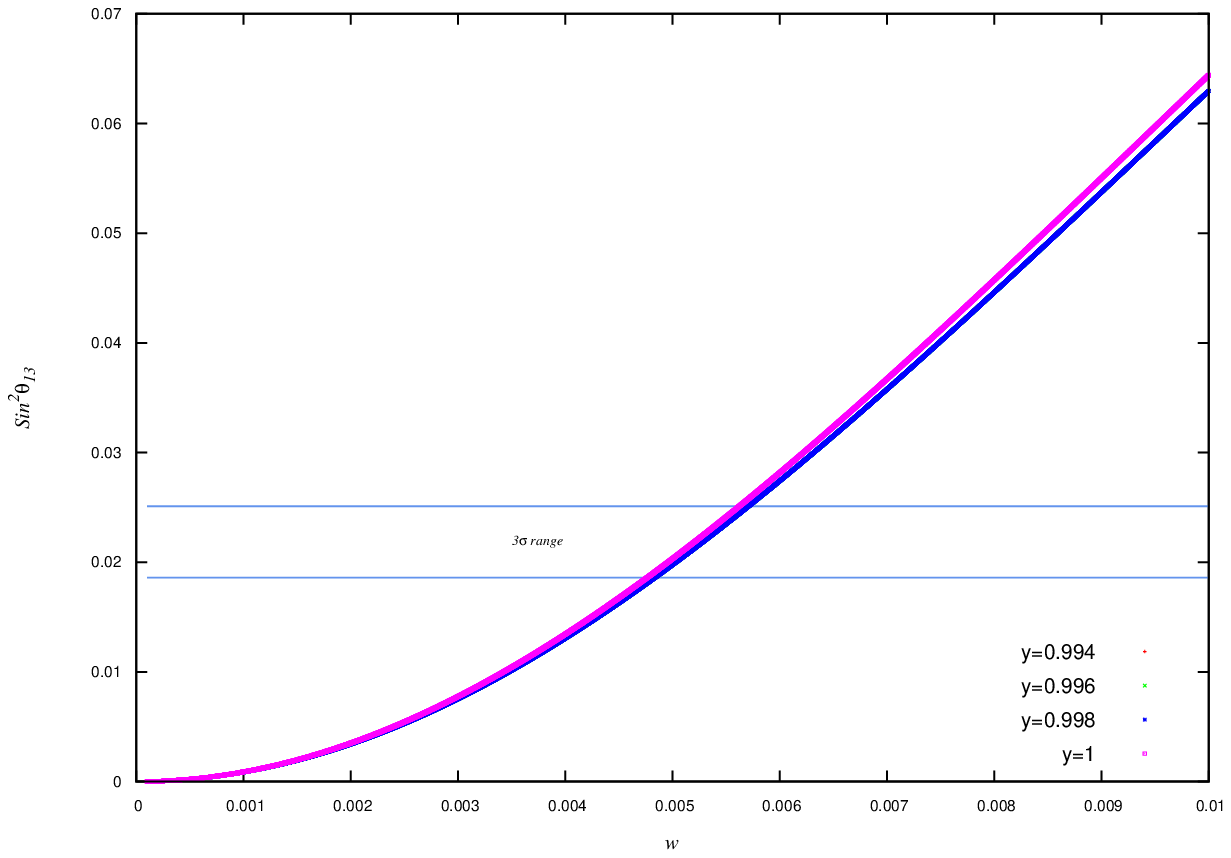}
 \caption{Generation of non-zero $sin^2\theta_{13}$, varying the type II strength using lower bound of $3\sigma$ deviations. }
 \label{fig7}
 \end{figure*}

\begin{figure*}
  \centering
  \includegraphics[width=10cm,height=8cm]{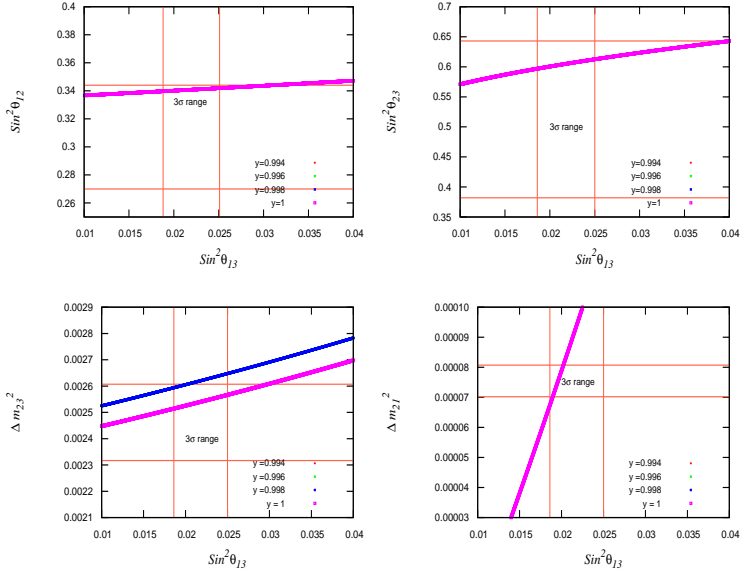}
  \caption{Variation of $sin^2\theta_{12}$, $sin^2\theta_{23}$ ,$\Delta m_{23}^2$ and $\Delta m_{21}^2$ with $sin^2\theta_{13}$
  with best fit value.}
  \label{fig8}
  \end{figure*}

\begin{figure*}
  \centering
  \includegraphics[width=10cm,height=8cm]{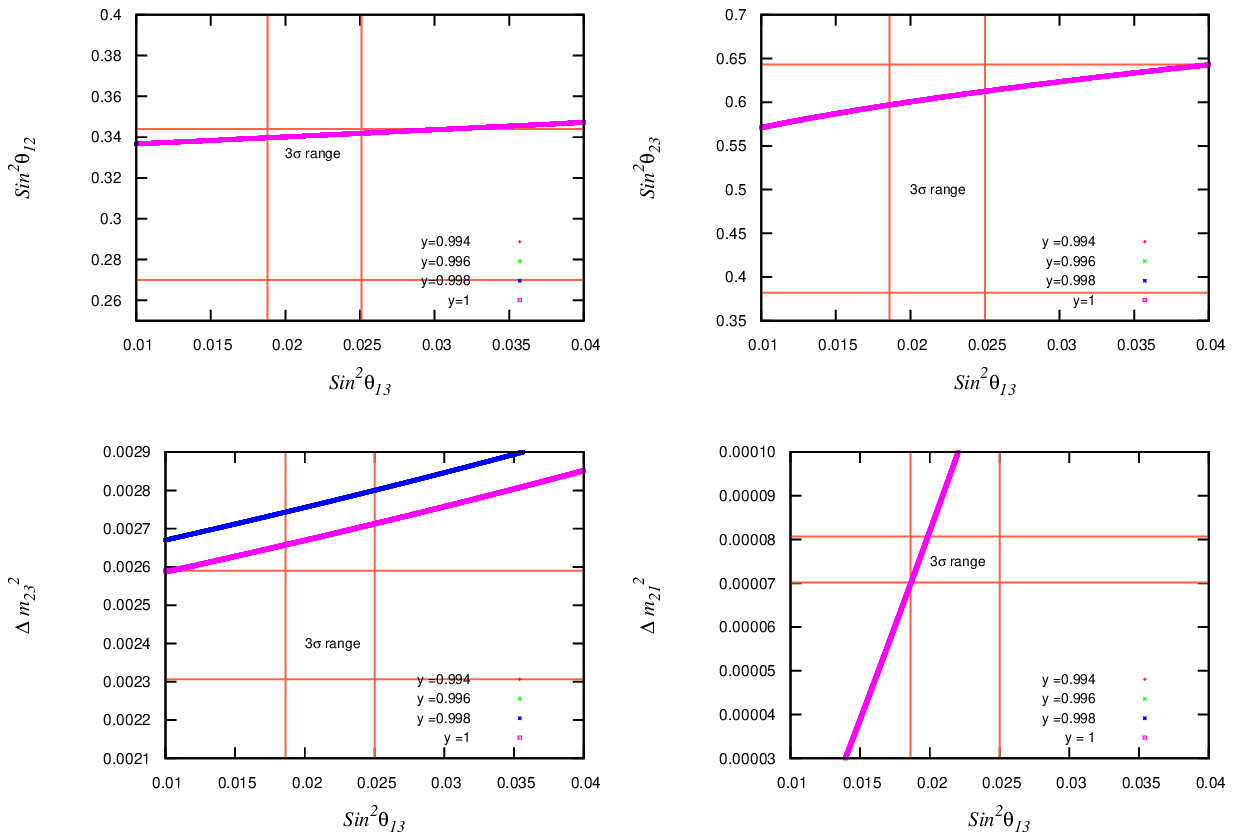}
  \caption{Variation of $sin^2\theta_{12}$, $sin^2\theta_{23}$ ,$\Delta m_{23}^2$ and $\Delta m_{21}^2$ with $sin^2\theta_{13}$
  with upper bound of $3\sigma$ deviation.}
  \label{fig9}
  \end{figure*}

\begin{figure*}
  \centering
  \includegraphics[width=10cm,height=8cm]{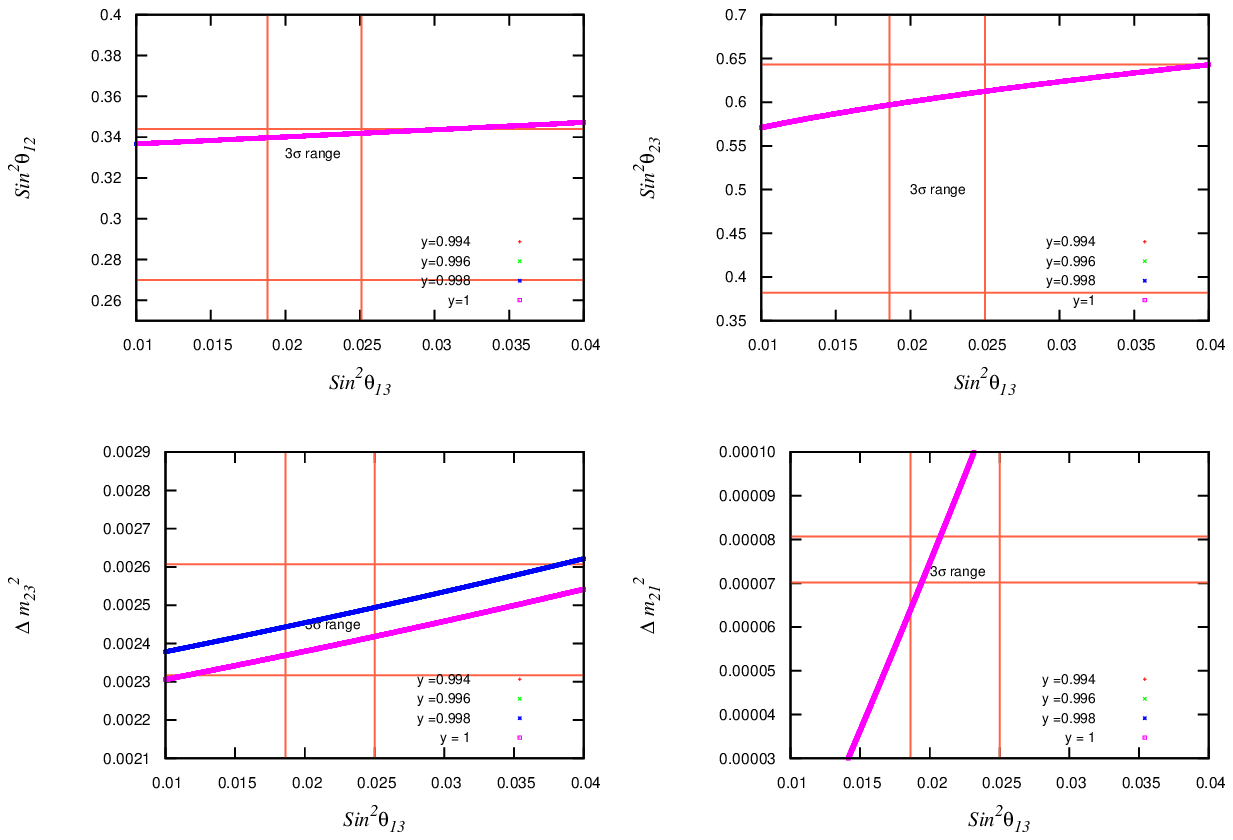}
  \caption{Variation of $sin^2\theta_{12}$, $sin^2\theta_{23}$ ,$\Delta m_{23}^2$ and $\Delta m_{21}^2$ with $sin^2\theta_{13}$ 
  with lower bound of $3\sigma$ deviation.}
  \label{fig10}
  \end{figure*}

\begin{figure*}
  \centering
  \includegraphics[width=10cm,height=8cm]{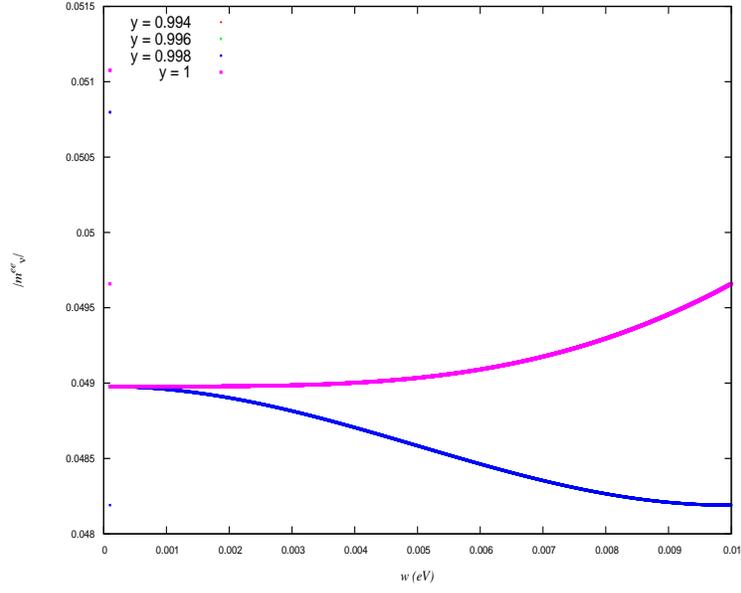}
  \caption{Variation of effective mass $m^{ee}_{\nu}$ with type II seesaw strength using bfp.}
  \label{fig11}
  \end{figure*}

\begin{figure*}
  \centering
  \includegraphics[width=10cm,height=8cm]{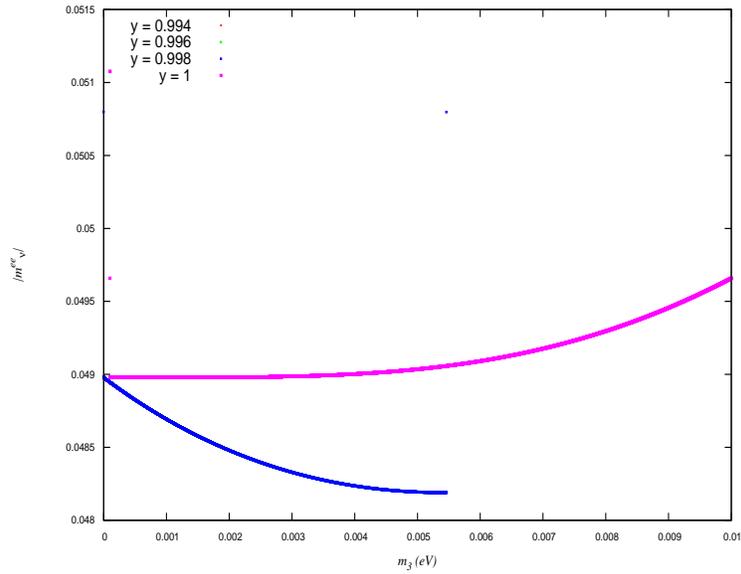}
  \caption{Variation of effective mass $m^{ee}_{\nu}$ with the lightest neutrino mass using bfp.}
  \label{fig12}
  \end{figure*}

\begin{figure*}
  \centering
  \includegraphics[width=10cm,height=8cm]{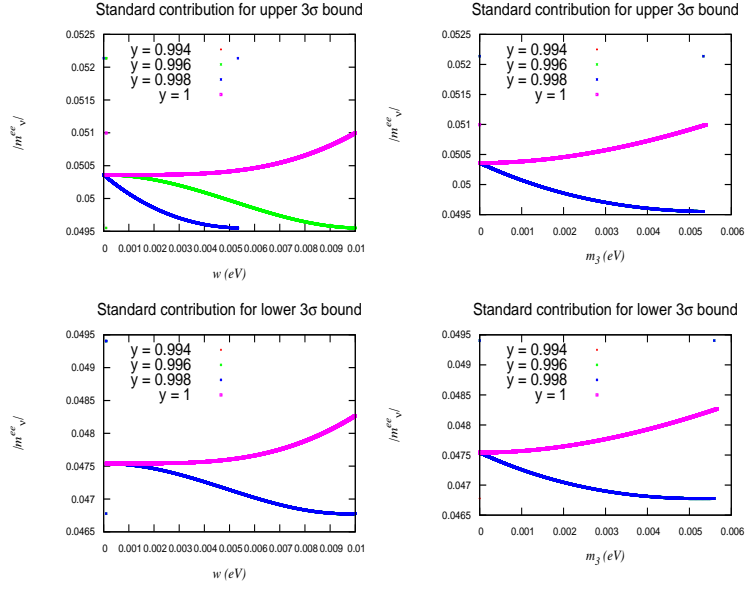}
  \caption{Variation of effective mass $m^{ee}_{\nu}$ with type II seesaw strength and $m_{3}$ for upper and lower $3\sigma$ bounds .}
  \label{fig13}
  \end{figure*}

   The scattering cross section of the decay channel described by figure \ref{fig3} to various SM fermions has been calculated. They are found
   to have an order of $10^{-60 }cm^{2}$ / $10^{-42} GeV^{-2}$ which is much smaller than the cross section which has been achieved for the t-channel
   contribution(of the order of $10^{-44} cm^{2}$). They will have little contribution (can be neglected therefore) to the relic abundance
   of the potential DM candidate. We have already noticed that for obtaining the observed $\Omega$ we need to fix the Yukawa coupling. Fixing 
    the Yukawa coupling as varying from 0.99 to 1, varying $m_{DM}$ from 30 to 60 GeV and varying $M_{R}$ from 
    120 to 167 GeV, we study the order of relic abundance. We fit the values of oscillation parameters using
    recent cosmological constraints for inverted mass ordering. 
    We compute all the oscillation parameters also by varying
    the type II seesaw strength. Variation of type II seesaw strength with the non-vanishing $\theta_{13}$,  has been 
    shown in figure \ref{fig5}, figure \ref{fig6}, figure \ref{fig7}. 
    The production of other oscillation parameters, e.g. the two mixing angles and two masses
    squared splitting as a function of nonzero $\theta_{13}$ has been shown in the figure \ref{fig8}, figure \ref{fig9} and figure \ref{fig10} 
    for different values of Yukawa coupling. The sum of absolute masses has
    also been calculated to see whether it satisfies the Planck upper bound or not. Seeing that, the sum of absolute neutrino masses can
    give some clue on neutrinoless double beta decay, a little study has been performed to check 
    whether the presented model is able to contribute to the $0\nu\beta\beta$ physics. In figure \ref{fig11} we plot for 
    the contribution of the effective mass to $0\nu\beta\beta$ decay due to light neutrino
    exchanges for standard contribution showing the variation of effective mass with the type II seesaw strength. Figure \ref{fig12} displays the variation of $m^{ee}_{\nu}$ 
   with the lightest neutrino  mass, in our model $m_{3}$. In figure \ref{fig13} we present the variation of effective mass with $m_{lightest}$ and type
   II seesaw strength taking the upper and lower bound of $3\sigma$ deviation. Since the presented model only presents a hierarchy
   of inverted kind the lowest mass range has
   been selected which is resulted from the perturbation. The variation in $m^{ee}_{\nu}$ for non-standard contribution with different $y$ values have been
   checked and found to be in agreement
    with the experimental bounds. The effective mass for non-standard contribution has been obtained around $0.0489$ almost for all the values of Yukawa
    couplings chosen for the analysis.
    It is worth noting that the variation in Yukawa coupling leaves trivial impacts on $m^{ee}_{\nu}$ for non-standard contribution. 
    For showing the variation of $m^{ee}_{\nu}$ with $m_{3}$, we choose those values of $m_{3}$ obtained as a result of adding the type II seesaw strength.
   
  The following observations have been made from the results and analysis.
\begin{itemize}
\item The relic abundance has been found to match the value shown by PLANCK 2013 data, for a choice of Yukawa coupling 
ranging from $0.99$ to $1$ provided the Relic mass is fixed at 50 GeV keeping the mediator mass at a range from 153 to 154 GeV. A detailed analysis of 
the choice of Yukawa coupling, the Relic mass($m_{\chi}$) and the mediator mass($m_{\psi}$) for this particular model has been presented in the 
table \ref{tab10}.
\item  The oscillation parameters are near to reach only when the Yukawa coupling is varied from 0.994 to 1 and as a further increase/decrease of 
the Yukawa coupling does not yield good neutrino phenomenology we have considered those corresponding values of relic abundance obtained for
Yukawa coupling ranging from 0.994 to 1.
\item  It has been noticed that the proposed model evidences correct neutrino phenomenology using the best fit and lower $3\sigma$ 
 bound in case of inverted hierarchy mass pattern only. All the oscillation parameters have been seen to come inside the frame while taking
 the best fit and lower $3\sigma$ 
 bound.
\item The non-zero value of $\theta_{13}$ has been found to be consistent with the variation of type II seesaw strength.
\item Both the standard and new physics contribution to $0\nu\beta\beta$ decay in the allowed hierarchy is obtained in the vicinity of experimental
results [GERDA].
\end{itemize}

%   
% %   

 \clearpage        
 \clearpage
\section{\label{sec:level8} Conclusion} 
An $A_{4}$ based IH neutrino mass model originating from both Inverse and type II seesaw have been studied. Here ISS is implemented as
 a leading order contribution to the light neutrino mass matrix yielding zero reactor mixing and $m_{3} = 0$. Then the type II seesaw
  has been used in order to produce non-Zero reactor mixing angle, which later on produces $m_{3} \neq 0$ keeping the hierarchy as inverted only.
  We have studied the possibility of
  having a common parameter space where both the Neutrino oscillation parameters in the $3\sigma$ range and
DM relic abundance has a better reach. With a proper choice of Yukawa coupling($y$), right handed neutrino (mediator particle) mass ($m_{\psi}$) ,
and complex scalar (potential DM candidate) mass ($m_{\chi}$) the variation in relic abundance as a function of Yukawa coupling has been 
shown. For a choice of Yukawa coupling between 0.994 to 0.9964, $m_{DM}$ around 50 GeV, the mediator mass needs to fall around
153 GeV to match the correct relic abundance. The same Yukawa coupling has got a key role in generating 
the Neutrino oscillation parameters as well. We have studied the prospect of producing non-zero $\theta_{13}$ by introducing a perturbation to
the light neutrino mass matrix using type II seesaw within the $A_{4}$ model. We have also determined the 
strength of the type II seesaw term which is responsible for the generation of non-zero $\theta_{13}$ in the correct $3\sigma$
range. We have also checked whether the proposed model can project about neutrinoless double beta decay or not. In context to the presented model we have 
found a wide range of parameter space where one may have a better reach for both neutrino and dark matter sector as well. This model may have relevance in
studying baryon asymmetry of the universe, which we leave for future study.
\begin{table}[htb]   
    \centering
     \begin{tabular}{|c|c|c|c|}
    \hline   $m_{\chi}$ &  $m_{\psi}$ & $y$ & $\Omega$  \\ 
    \hline $30$ GeV & $(121 - 122)$ GeV & $(0.99 - 1)$ & $ \checkmark$ \\ 
    \hline $40$ GeV &  $(139)$ GeV & $(0.99 - 1)$ & $ \checkmark$ \\ 
    \hline $50$ GeV & $(153 - 154)$ GeV & $(0.99 - 1)$ & $\checkmark$ \\ 
    \hline $60$ GeV & $(166 - 167)$ GeV & $(0.99 - 1)$ & $\checkmark$ \\
    \hline
     \end{tabular}
  \caption{Comparison of relic abundance $\Omega$ with various choices of Yukawa couplings,
  DM mass, RH neutrino mass} \label{tab10}      
     \end{table}
     
 \clearpage    
    
     \begin{table}[h!]
    \centering
     \begin{tabular}{|c|c|c|c|c|c|c|}
    \hline     $3\sigma$ ranges                  &  $ \theta_{13} $ & $\theta_{12} $ & $\theta_{23} $ & $\Delta m_{21}^2 $ & 
     $\Delta m_{23}^2$ & $\Sigma\mod{m_{i}}$\\ 
    \hline  bfp                   & $\checkmark$  & $\checkmark$ & $\checkmark$ & $\checkmark$ & $\checkmark$ & $\checkmark$ \\
    \hline lower bound  & $\checkmark$  & $\checkmark$ & $\checkmark$ & $\checkmark$ & $\checkmark$ & $\checkmark$ \\
    \hline  upper bound & $\checkmark$  & $\checkmark$ & $\checkmark$ & $\checkmark$ & $\times$ & $\checkmark$ \\
    \hline
    \end{tabular}
  \caption{Summery of results obtained from various allowed mass schemes.} \label{tab11}      
   \end{table}
   
 \textbf{ ACKNOWLEDGMENTS :}

 It is our pleasure to acknowledge Dr. Debasish Borah for useful discussions in addressing the stability of the Dark Matter.
 The authors gratefully acknowledge the anonymous referee for useful comments and suggestions which helped us to improve the manuscript. The work of 
 MKD is partially supported by the grant no. 42-790/2013(SR) from UGC, Govt. of India.

\end{document}